\documentclass[
final            
  ]
  {aipproc}

\layoutstyle{6x9}

\usepackage{amsmath}                    
\usepackage{amssymb}                    
\usepackage{verbatim}
\usepackage{bm}
\newcommand{\w}[1]{\bm{#1}} 

\begin{document}

\title{New results for Petrov type D \\
	pure radiation fields}


\author{Liselotte De Groote}{
  address={Ghent University, Department of Mathematical Analysis\\
  Galglaan 2, 9000 Ghent, Belgium}
}

\author{Norbert Van den Bergh}{
  address={Ghent University, Department of Mathematical Analysis\\
  Galglaan 2, 9000 Ghent, Belgium}
}

\begin{abstract}
	We present a new family of Petrov type D pure radiation spacetimes with a shear-free, non-diverging geodesic principal null congruence.
\end{abstract}

\maketitle


\section{Introduction}
	
	In 1990 Wils\cite{Wils} showed that aligned (and therefore geodesic and shear-free) pure radiation fields of Petrov type D, with or without cosmological constant, are necessarily non-twisting. The diverging solutions then belong to the Robinson-Trautman class and are all explicitly known\cite{Kramer}. In the non-diverging case, however, only a few type D examples are available\cite{WilsVdB}; here we present new solutions belonging to the latter class.

A pure radiation field has an energy-momentum tensor of the form $T_{ab} = \phi k^a k^b$ with $k_a k^a = 0$. Solutions are said to belong to Kundt's class, if the spacetime also admits a null congruence, generated by a vectorfield $\tilde{k}$, which is non-diverging (i.e.~which is both non-expanding and twist-free and therefore also geodesic).
	For vacuum and physically reasonable matter content, the Goldberg-Sachs theorem implies that the Kundt spacetimes must be algebraically special (type II, D, III or N or conformally flat), with $\tilde{k}$ being the repeated principal null direction of the Weyl tensor. In the case of a pure radiation field one can furthermore show\cite{Kramer}, making use of the null energy condition, that $k$ and $\tilde{k}$ are aligned
and that the associated null congruence is shear-free. It is known that, in the physically important case of a Petrov type D spacetime, no pure radiation field can be a (null) Maxwell field\cite{DebeverVdBLeRoy}, or a (null) neutrino or scalar field\cite{WilsVdB}.

As shown by Kundt\cite{Kundt} in 1961 the line-elements admitting a geodesic, shear-free and non-diverging null congruence can all be expressed in the form:
	\begin{equation}
		\label{dskundt}
		\mathrm{d} s^2 = 2 P ^{-2} \mathrm{d}\zeta \mathrm{d}\overline{\zeta} - 2 \mathrm{d}u (\mathrm{d}\textrm{v} + H\mathrm{d}u + W\mathrm{d}\zeta + \overline{W} \mathrm{d} \overline{\zeta}),
	\end{equation}
	in which $P$ is a real function of $(\zeta, \overline{\zeta},u)$ and $H, W$ are respectively real and complex functions of $(\zeta,\overline{\zeta},u,\textrm{v})$, to be determined by appropriate field equations.

\par

The vacuum solutions of this type have been known for a very long time\cite{Kinnersley}. A procedure (based on Theorem 31.1 in~\cite{Kramer}) is also available allowing one to generate non-vacuum solutions of Kundt's class, from vacuum solutions. However, in this way the Petrov type of the metric generally will be changed from D to II. Insisting that the Petrov type does not change, constrains the function $H_0$ of the `background-metric' and it was not certain whether one could generate all pure radiation solutions in this way. This technique was used in \cite{WilsVdB}, where the authors managed to construct a family of type D pure radiation fields and where they conjectured that these solutions were the only aligned type D pure radiation fields of Kundt's class. Below we show that the solutions obtained in \cite{WilsVdB} only cover a small part of the entire family. In a follow-up paper\cite{DeGrooteVdB3} we will show that the solutions presented here actually \emph{exhaust} the full aligned Petrov type D pure radiation class.
	
\section{Kundt type D pure radiation fields}
	The most obvious way to find all type D pure radiation fields of Kundt's class, is to start from the general Kundt metric (\ref{dskundt}), express that the solutions we are looking for are of Petrov type D, and to make sure that the field equations for pure radiation are satisfied. This however introduces a hard to solve system of non-linear conditions. We therefore prefer to start from the general Newman Penrose equations and to extract from these all possible invariant information, before introducing any coordinates.
	\par
	First we introduce the basic assumptions expressing that the null congruence tangent to $k$ is geodesic, shear-free and non-diverging, $\kappa = \sigma = \rho = 0$.
Next we use the type D condition in order to choose the null tetrad such that $\Psi_2$ and $\Phi_{22}$ are the only non-vanishing components of respectively the Weyl tensor and of the traceless part of the Ricci tensor.
From the Bianchi identities it then follows that the spin-coefficient $\lambda =0$, while appropriate boosts and rotations allow one to put $\epsilon = 0$ and $\tau -\overline{\alpha} - \beta =0$. Next a rather technical proof shows that the spin-coefficient $\pi$ has to be real. This divides the solutions into two cases: one in which $\pi=0$ and one in which $\pi\neq 0$.

\subsection{The case $\pi = 0$}
	If $\pi = 0$ we obtain the following explicit expressions for the spin-coefficient $\mu$ and curvature components: $\mu =0$, $\Psi_2 = -\frac{R}{12}$, $\Phi_{22} = -4 \alpha \nu$.
We also obtain the following total derivatives ($\w{\omega}^i$ representing the basis one-forms):
	\begin{align*}
		\mathrm{d} \alpha &= (2 \alpha^2 + \frac{R}{16} )(\w{\omega}^1 + \w{\omega}^2) + \Delta \alpha \w{\omega}^3, \\
		\mathrm{d} \gamma &= \delta \gamma \w{\omega}^1 + \overline{\delta} \gamma \w{\omega}^2 + \Delta \gamma \w{\omega}^3 - \frac{R}{8} \w{\omega}^4, \\
		\mathrm{d} \nu &= -2\alpha \nu (\w{\omega}^1 + \w{\omega}^2) + \Delta \nu \w{\omega}^3.
	\end{align*}
	The remaining equations to be satisfied are then given by:
	\begin{align*}
		\delta \overline{\gamma} + \delta \gamma &= 0, \\
		\Delta \alpha - \overline{\delta} \gamma &= \alpha (\overline{\gamma} - \gamma), \\
		\overline{\delta}\gamma - \delta \overline{\gamma} &= 2 \alpha (\gamma - \overline{\gamma}).
	\end{align*}
	From the expression for $\Phi_{22}$ and the total derivatives of $\alpha$ and $\nu$, one can see that both $\alpha$ and $\nu$ can not be constants. This allows one, after a few calculations, to rewrite the basis one-forms as follows:
	\begin{align*}
		\w{\omega}^1 &= \frac{2\mathrm{i}\alpha \nu B  + \Delta \nu }{4 \alpha \nu} \mathrm{d}u - \frac{1}{4 \alpha \nu} \mathrm{d} \nu + \frac{1}{2} \mathrm{i} C \mathrm{d}x, \\
		\w{\omega}^3 &= \mathrm{d}u ,\ \w{\omega}^4 = \mathrm{d}\textrm{v} + H \mathrm{d}u,
	\end{align*}
	where $B$ and $C$ are real functions of $(u, \nu, x)$ and $H$ is a real function of $(u, \textrm{v}, \nu, x)$. The remaining system of equations can be fully integrated. Two subclasses have to be distinguished (in both cases $b, m$ and $n$ are arbitrary functions of $u$, while $a$ is a parameter related to the Ricci scalar by $R=-32 a^2$):
	\subsubsection{$\Delta \alpha = 0$}
		If $\Delta \alpha = 0$ we obtain the basis one-forms:
		\begin{align}
			\w{\omega}^1 &= -\frac{1}{4}\frac{(xy+2\mathrm{i})^2 \mathrm{i}b}{y}\mathrm{d}u+\frac{\mathrm{i}y}{8 a}\mathrm{d}x + \frac{1}{4 a y}\mathrm{d}y,  \nonumber \\
			\w{\omega}^3 &= \mathrm{d} u, \ \w{\omega}^4 = \mathrm{d}\textrm{v} + \frac{1}{2} (8 a^2 \textrm{v}^2 -my + 2 n) \mathrm{d}u,
		\end{align}
		resulting in $\alpha=a$ and $\nu = a m y$.
	\subsubsection{$\Delta \alpha \neq 0$}
		If $\Delta \alpha \neq 0$ the basis one-forms reduce to:
		\begin{align}
			\w{\omega}^1 &= -\frac{1}{4} (\cosh x-\mathrm{i}\,\sinh x\,\sinh y)b \mathrm{d}u + \frac{\mathrm{i} \cosh y}{4 a}\mathrm{d}x + \frac{1}{4 a}\mathrm{d}y,  \nonumber \\
			\w{\omega}^3 &= \mathrm{d} u,\ \w{\omega}^4 = \mathrm{d} \textrm{v} + \frac{1}{2}(8 a^2 \textrm{v}^2-m \sinh y + 2 n ) \mathrm{d}u,
		\end{align}
		resulting in $\alpha= a \tanh y$ and $\nu=a m \cosh y$.

\subsection{The case $\pi \neq 0$}
	In this case we find explicit expressions for the spin-coefficients $\alpha, \beta, \mu$ and for $\Phi_{22}$:
	\begin{align*}
		\alpha &= \frac{1}{4} \frac{\Psi_2}{\pi} + \frac{1}{2}\frac{L}{\pi} - \frac{1}{2} \pi & \beta = -\frac{1}{4} \frac{\Psi_2}{\pi} - \frac{1}{2}\frac{L}{\pi} - \frac{1}{2} \pi \\
		\mu &= 0   & \Phi_{22} = -\frac{\nu (\Psi_2 + 2L)}{\pi}.
	\end{align*}
	We also find the total derivatives:
	\begin{align*}
		\mathrm{d} \gamma &= \Delta \gamma \w{\omega}^3 + (\Psi_2 -\pi^2 - L) \w{\omega}^4, \\
		\mathrm{d} \nu &= - \frac{\nu(\Psi_2 + 2L)}{2 \pi} (\w{\omega}^1 + \w{\omega}^2) + \Delta \nu \w{\omega}^3, \\
		\mathrm{d} \pi &= -(\pi^2 + \frac{\Psi_2}{2} + L) (\w{\omega}^1 + \w{\omega}^2), \\
		\mathrm{d} \Psi_2 &= -3\pi \Psi_2 (\w{\omega}^1 + \w{\omega}^2),
	\end{align*}	
	from which we see that $\Psi_2 = \Psi_2(\pi)$, in particular:
	\begin{equation}
		-\Psi_2 + \pi^2 + L = k (c\Psi_2)^{2/3}, \qquad k,c \; \mathrm{constants}.
	\end{equation}
	Note that  $\pi$ is not a constant, otherwise $\Psi_2$ would be constant and hence $\pi=0$. This shows that we can use $\pi$ (or $\Psi_2$) as a coordinate, but we prefer to write $\pi = \pi(z)$, $\Psi_2 = \Psi_2(z)$ etc.
	\par
	In this case we can write the basis one-forms as follows:
	\begin{align}
		\w{\omega}^1 &= \frac{1}{2 p} \mathrm{d} z + \mathrm{i} p \mathrm{d}y,  \nonumber \\
		\w{\omega}^3 &= \mathrm{d} u, \ \w{\omega}^4 = \mathrm{d} \textrm{v} - 2\frac{\textrm{v}}{z} \mathrm{d}z - \frac{k c^2 \textrm{v}^2-m^2z^3-n z^4}{z^2}\mathrm{d}u,
	\end{align}
	with arbitrary functions $m$ and $n$ of $u$, and $ p = \sqrt{c^2/z+kc^2-Lz^2}$.

This results in $\pi=-p/z$, $\Psi_2=c^2/z^{3}$ and yields the line-element:
	\begin{equation*}
		\mathrm{d}s^2 = 2\frac{m^2z^3 + n z^4-kc^2\textrm{v}^2}{z^2}\mathrm{d}u^2 + 2\mathrm{d}u\mathrm{d}\textrm{v}-4\frac{v}{z}\mathrm{d}u \mathrm{d}z -2p^2\mathrm{d}y^2 - \frac{1}{2p^2}\mathrm{d}z^2,
	\end{equation*}
	which for $k = 1$ and $m = 0$ reduces to the metric of \cite{WilsVdB}.
	

\bibliographystyle{aipproc}   


\begin{thebibliography}{7}
\expandafter\ifx\csname natexlab\endcsname\relax\def\natexlab#1{#1}\fi
\providecommand{\enquote}[1]{``#1''}
\expandafter\ifx\csname url\endcsname\relax
  \def\url#1{\texttt{#1}}\fi
\expandafter\ifx\csname urlprefix\endcsname\relax\def\urlprefix{URL }\fi
\providecommand{\eprint}[2][]{\url{#2}}

\bibitem[Wils(1990)]{Wils}
P.~Wils, \emph{Class. Quantum Grav.} \textbf{7}, 1905 (1990).

\bibitem[Kramer et~al.(2003)]{Kramer}
D.~Kramer, H.~Stephani, M.~A.~H. MacCallum, C.~Hoenselaers and E.~Herlt,
  \emph{Exact solutions of Einstein's field equations}, Cambridge University
  Press, 2003.

\bibitem[Wils and {Van den Bergh}(1990)]{WilsVdB}
P.~Wils and N.~{Van den Bergh}, \emph{Class. Quantum Grav.} \textbf{7}, 577
  (1990).

\bibitem[{Van den Bergh}(1989)]{DebeverVdBLeRoy}
N.~{Van den Bergh}, \emph{Class. Quantum Grav.} \textbf{6}, 1373 (1989).

\bibitem[Kundt(1961)]{Kundt}
W.~Kundt, \emph{Z. Phys.} \textbf{163}, 77 (1961).

\bibitem[Kinnersley(1969)]{Kinnersley}
W.~Kinnersley, \emph{J. Math. Phys.} \textbf{10}, 1195 (1969).

\bibitem[{De Groote} and {Van den Bergh}(2009)]{DeGrooteVdB3}
L.~{De Groote} and N.~{Van den Bergh}, \emph{to be published}  (2009).

\end{thebibliography}

\end{document}